%% file: main.tex
\pdfoutput=1
\documentclass[sigconf]{acmart}

\usepackage{subcaption}
\usepackage[subtle]{savetrees}
\usepackage{multirow}
\usepackage{microtype}
\usepackage{multicol}
\usepackage{tabularx}
\usepackage{graphics}
\usepackage{adjustbox}
\usepackage{nicefrac}
\usepackage{booktabs}

\usepackage{setspace}
\usepackage{enumitem,kantlipsum}
\usepackage{pgfplots}
\usetikzlibrary{pgfplots.groupplots}
\pgfplotsset{compat=1.11}
\usepgfplotslibrary{ternary}
\usepackage{tikz}
\usepackage{tkz-graph}
\usetikzlibrary{positioning,automata}
\usetikzlibrary{calc,spy,shapes}
\usetikzlibrary{arrows,decorations.pathmorphing,fit,positioning,patterns}
\pgfplotsset{compat=newest}

\usepackage{xspace} % insert space when needed.
\newcommand{\acg}{ACG\xspace}

\def\:{\hskip0pt} % \: for hyphenation xxx\:---\:xxx 
\newcommand{\mypar}[1]{\vspace*{-0.1ex}\medskip\noindent\textbf{#1}~}
\newcounter{todocnt}

\DeclareFontFamily{U}{mathb}{\hyphenchar\font45}
\DeclareFontShape{U}{mathb}{m}{n}{
<-6> mathb5 <6-7> mathb6 <7-8> mathb7
<8-9> mathb8 <9-10> mathb9
<10-12> mathb10 <12-> mathb12
}{}
\DeclareSymbolFont{mathb}{U}{mathb}{m}{n}

\DeclareMathSymbol{\smalltriangleup} {2}{mathb}{"98}% name to be checked
\DeclareMathSymbol{\smalltriangledown} {2}{mathb}{"99}% name to be checked
\DeclareMathSymbol{\smalltriangleleft} {2}{mathb}{"9A}% name to be checked
\DeclareMathSymbol{\smalltriangleright}{2}{mathb}{"9B}% name to be checked
\DeclareMathSymbol{\blacktriangleup} {2}{mathb}{"9C}% name to be checked
\DeclareMathSymbol{\blacktriangledown} {2}{mathb}{"9D}% name to be checked
\DeclareMathSymbol{\blacktriangleleft} {2}{mathb}{"9E}% name to be checked
\DeclareMathSymbol{\blacktriangleright}{2}{mathb}{"9F}% name to be checked

\renewcommand\vec[1]{\overrightarrow{#1}}
\newcommand\cev[1]{\overleftarrow{#1}}

\copyrightyear{2017}
\acmYear{2017}
\setcopyright{acmlicensed}
\acmConference{CIKM'17}{}{November 6--10, 2017, Singapore.}
\acmISBN{ISBN 978-1-4503-4918-5/17/11}
\acmPrice{$15.00$} \acmDOI{https://doi.org/10.1145/3132847.3133010}

\begin{document}

\title{Learning to Attend, Copy, and Generate \\ for Session-Based Query Suggestion}
\renewcommand{\shorttitle}{Learning to Attend, Copy, and Generate for Session-Based Query Suggestion}

\author{Mostafa Dehghani}
\authornote{Work done while interning at Google Research.}
\affiliation{%
  \institution{University of Amsterdam}
}
\email{dehghani@uva.nl}

\author{Sascha Rothe}
\affiliation{%
  \institution{Google Research}
  }
\email{rothe@google.com}

\author{Enrique Alfonseca}
\affiliation{%
  \institution{Google Research}
  }
\email{ealfonseca@google.com}

\author{Pascal Fleury}
\affiliation{%
  \institution{Google Research}
  }
\email{fleury@google.com}

\renewcommand{\shortauthors}{M. Dehghani et al.}

\newcommand{\maingoal}{}
\newcommand{\rqone}{}
\newcommand{\rqtwo}{}
\newcommand{\rqthree}{}

\begin{abstract}
Users try to articulate their complex information needs during search sessions by reformulating their queries.
To make this process more effective, search engines provide related queries to help users in specifying the information need in their search process.
In this paper we propose a customized sequence-to-sequence model for session-based query suggestion.
In our model, we employ a query-aware attention mechanism to capture the structure of the session context. 
This enables us to control the scope of the session from which we infer the suggested next query, which helps not only handle the noisy data but also automatically detect session boundaries.
Furthermore we observe that, based on the user query reformulation behavior, within a single session a large portion of query terms is retained from the previously submitted queries and consists of mostly infrequent or unseen terms that are usually not included in the vocabulary. 
We therefore empower the decoder of our model to access the source words from the session context during decoding by incorporating a copy mechanism.
Moreover, we propose evaluation metrics to assess the quality of the generative models for query suggestion. We conduct an extensive set of experiments and analysis.
The results suggest that our model outperforms the baselines both in terms of the generating queries and scoring candidate queries for the task of query suggestion.
\end{abstract}



% We no longer use \terms command

\maketitle
% \vfill
% \pagebreak

\section{Introduction}
Users interact with search engines during search sessions and try to direct their search by submitting a sequence of queries. 
Based on these interactions, search engines provide a prominent feature, in which they assist their users to formulate their queries to better represent their intent during Web search by providing suggestions for the next query.

Query suggestion might address the need for disambiguation of the user queries to make the direction of the search more clear for both, the user and the search engine. 
It might help users by providing a precise and succinct query when they are not familiar with the specific terminology or when they lack understanding of the internal vocabulary and structures in order to be able to formulate an effective query. 
It has been shown that in general, query suggestion accelerates search satisfaction by either diving deeper into the current search direction or by moving to a different aspect of a search task~\citep{Ozertem:2012,Shokouhi:2013}.

There has been a lot of research on the task of query suggestion and similar tasks like query auto-completion.
A large body of methods leverages the idea of the ``wisdom of crowds'' by analyzing the search logs to use either query co-occurrences~\citep{Huang:2003,Ozertem:2012} in the search logs, or document clicks information~\citep{Mei:2008,Baeza:2004,Sadikov:2010}. 
However, co-occurrence based models suffer from data sparsity and lack of coverage for rare or unseen queries. 
On the other hand, considering the previously issued queries in the session, i.e context queries, and their order as a sequence of attempts for finding relevant information is of crucial for providing an effective suggestion. 
Dealing with these highly diverse sessions makes using co-occurrence based model almost impossible~\citep{Sordoni:2015,Cao:2008,He:2009}.

\begin{figure*}[!t]%
    \centering
    \includegraphics[height=3.5cm]{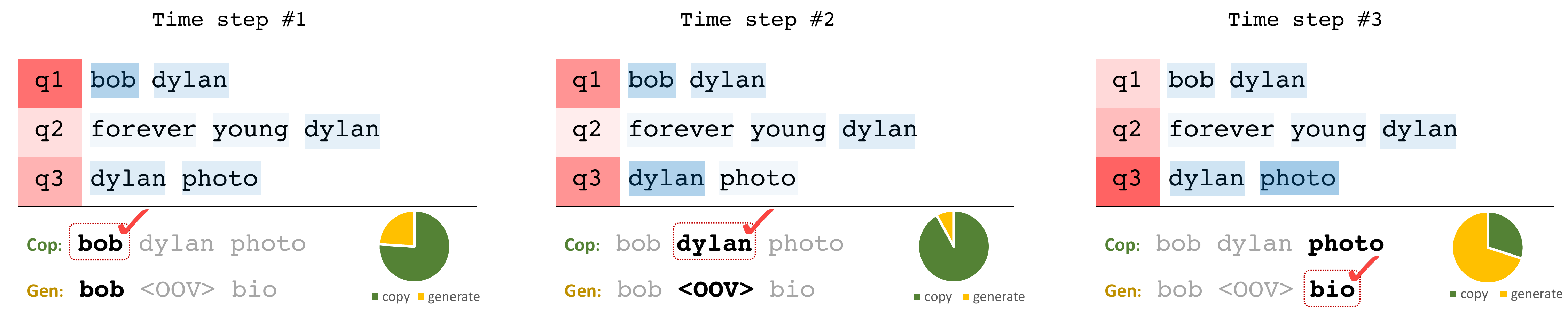}
    \caption{\label{fig:example}Example of generating a suggestion query given the previous queries in the session. The suggestion query is generated during three time steps.
    The heatmap indicates the attention, red for query-level attention and blue for word-level attention. The pie chart shows if the network decides to copy or to generate.}
 \vspace{-10pt}
\end{figure*}

Sessions are driven by query reformulations and users modifying existing queries in order to pursue new search results. 
Taking the structure of the context queries into account is important as query suggestion is well tightened to the understanding of query reformulation behaviors.
A good query suggestion system should be able to reproduce natural reformulation patterns from users.
There are several patterns in query reformulation like term addition, removal, and retention~\citep{Eickhoff:2014,Sloan:2015}. 
It has been shown that retained terms make up a large proportion of query reformulation in search sessions.
For example, an average of $62\%$ of the terms in a query are retained from their preceding queries~\citep{Sloan:2015}.
More than $39\%$ of the users repeat at least one term from their previous query~\citep{Jiang:2014}. 
On the other hand retained terms are clearly core terms indicating the user's information need, hence, they are usually discriminative terms and entities. 
Based on statistics from the AOL query log~\citep{Pass:2006}, more than 67\% of the retained terms in the sessions are from the bottom 10\% of terms ordered by their frequency.
%are from the bottom decile of terms ordered by their frequency.

The recent success of sequence-to-sequence (seq2seq) models in which recurrent neural networks (RNNs) both read and freely generate text makes it possible to generate the next query by reading the previously issued queries in the session~\citep{Sordoni:2015}. 
Although generic seq2seq models are promising in generating text, they have some shortcomings in the task of query suggestion.
The first problem of directly employing the generic seq2seq model for the task of query suggestion is that it considers the input data as a sequence of words, ignoring the query level information. To address this, \citet{Sordoni:2015} proposed a context-aware seq2seq model in which they use a hierarchical architecture to encode the previously issued queries in the session and generate the most likely sequence of words as the next query.
The second shortage of a generic word-based seq2seq model is that it's unable to deal with out-of-vocabulary words (OOV).
Besides, these models are less likely to generate terms with very low frequency~\citep{Fadaee:2017}.
This makes them unable to effectively model term retention, which is the most common reformulation patterns for the next query generation.

In this paper, we present an architecture that addresses these two issues in the context of session-based query suggestion.
We augment the standard seq2seq model with query-aware attention mechanism enabling the model to \textbf{\underline{A}ttend} to the promising scope of the session for generating the next query.
Furthermore, we incorporate the copy mechanism by adding a copier component which lets the decoder \textbf{\underline{C}opy} terms from the session context that improves the performance by modeling the term retention and handling of OOVs.
The model still has the ability to \textbf{\underline{G}enerate} new words through a generator component.
Our model, which we are going to call \textbf{\acg} in the rest of the paper, is trained in a multi-objective learning process.

Figure~\ref{fig:example} illustrates an example of the output of our model as the suggestion for the next query, given the previously submitted queries in a session. 
This example session is composed of three queries: \emph{bob dylan} $\longmapsto$ \emph{forever young dylan} $\longmapsto$ \emph{dylan photo}, which were submitted sequentially. 
Our model outputs the sequence of the words \emph{bob}, \emph{dylan}, and \emph{bio}. 
At each time step, the heatmap of the query level attention (red) and word level attention (blue) is illustrated.
Furthermore, the output of the copier, of the generator, and the probability of the network deciding to copy a term from the previous queries or to generate a new term is given for each time step. 
At time step \#1, the first query in the session and in this query, word \emph{bob} has the highest attention.
The outputs of both copier and generator are the same, but the network decides to copy the term \emph{bob} (probably from the first query).
At time step \#2, \emph{dylan} is an OOV. So the output of the generator is the $\langle OOV\rangle$ token and based on the learned attentions, the network decides to copy \emph{dylan} from queries in the session.
At time step \#3, the last query in the session and in this session term \emph{photo} has the highest attention, and the network decides to generate the new term \emph{bio}.

Besides proposing a seq2seq model which learns to effectively attend, copy and generate for the task of session-based query suggestion, we introduce new metrics for evaluating the output of generative models for the task of query suggestion.
We train and evaluate \acg on the AOL query log data and compare it to the state-of-the-art models both in terms of the ability to discriminate and the ability to generate.
The results suggest that \acg as a discriminative model is able to effectively score good candidates and as a generative model generates better queries compared to the baseline models.
In the following, we first explain our model in detail.
Then we describe the evaluation paradigm we use in this paper.
Afterward, we present our results followed by analysis and discussions.
In the end, we review some related work and conclude the paper.

\section{Our Proposed Model}
In this section, we first describe the seq2seq model with attention as one of the baselines and as the base model we build our model upon. 
We then introduce the general architecture of our model and explain the query-aware attention and copy mechanism as two main components employed in our model.

\begin{figure*}[!t]%
    \centering
    \includegraphics[height=6cm]{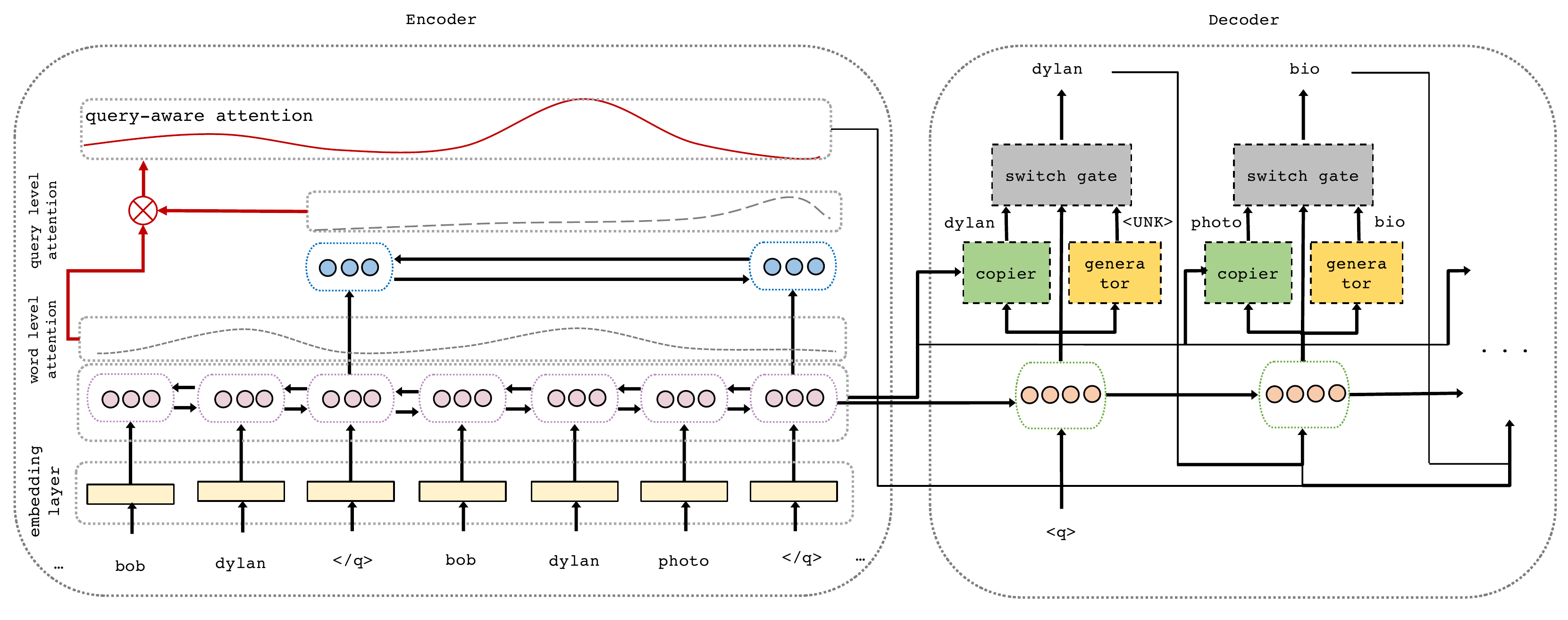}
    \caption{\label{fig:model}General schema of our proposed model for query suggestion.}
%  \vspace{-15pt}
\end{figure*}

\subsection{Seq2seq with Attention}
\label{sec:seq2seq_attention}
As the base model, we employ seq2seq (seq2seq) with attention, proposed by \citet{Bahdanau:2014} for the task of machine translation, which is able to jointly learn the translation and the alignment. 
In general, the seq2seq model is an RNN-based encoder-decoder in which the encoder learns the representation for the source sequence, and the decoder generates the target sequence.

The encoder is a bidirectional recurrent neural network (RNN) that reads the input sequence $X = [x_1, x_2,\ldots,x_n]$ in the left-to-right direction in the RNN forward pass.
It creates sequence of hidden states, $[\vec{h}_{1},\vec{h}_{2},\ldots,\vec{h}_{n}]$, where $\vec{h}_i = \text{RNN}(x_i,\vec{h}_{i-1})$ is a dynamic function for which we can use for example an LSTM~\citep{Hochreiter:1997} or a GRU~\citep{Cho:2014}.
The RNN backward pass reads $X$ in the reverse direction, i.e. $\cev{h}_i = \text{RNN}(x_i,\cev{h}_{i+1})$ resulting in a sequence of hidden states $[\cev{h}_{n},\cev{h}_{n-1},\ldots, \cev{h}_{1}]$.

The forward and backward states for each time step are concatenated to create the encoder hidden states $[h_{1},h_{2},\ldots,h_{n}]$, in which $h_{i}=[\vec{h}_{i};\cev{h}_{i}]$ is the encoded information from the corresponding token $x_{i}$ in the source sequence $X$, taking it's surrounding tokens into consideration.
The encoded hidden state are then summarized using a function $\phi$ to generate a fixed length vector which is called \emph{context vector}: $c = \phi(h_{1},h_{2},\ldots,h_{n})$. 

The decoder is a unidirectional RNN with hidden states $s_t$ which unfolds the context vector to the target sequence. During the decoding process, in order to predict the target token at each time step $t$, the attention mechanism is used to dynamically change the context vector $c_t$. 
To do so, for each token $x_i$ in the source sequence, we compute a weight $a_i$ that determines the contribution of $h_i$, to the $t^{\text{th}}$ token in the target sequence:
\begin{align}
l_{t,i} &= \eta(s_{t-1}, h_i) \\
a_{t,i} &= \frac{\exp(l_{t,i})}{\sum_j^n\exp(l_{t,j})} \label{worl_level_att}
\end{align}
where $\eta$ is a mapping (usually a multilayer perceptron (MLP) or a bilinear function), $s_{t-1}$ is the previous decoder's hidden state.
% and $y_{t-1}$ is the output of decoder at the previous time step.
This mapping will give us the logits $l$ which are normalized with a softmax layer to get the final attention weights $a$.
The context vector $c_t$ is calculated as follow:
\begin{equation}\label{context_vector}
c_t = \sum_{i}^{n}{a_{t,i} h_i}
\end{equation}

In the decoder, the hidden state is computed based on the previous state $s_{t-1}$, the context vector $c_t$, and the previous output token $y_{t-1}$:
\begin{equation}
s_t = \text{RNN}(s_{t-1}, y_{t-1}, c_t)
\end{equation}
An output projection layer is used to compute the conditional distribution over the vocabulary:
\begin{equation}\label{output_projection_1}
p(y_t|y_{<t}, X) = f_o(s_t)
\end{equation}

In the seq2seq model encoder and decoder are jointly trained to maximize the conditional log-likelihood of the $Y_i$, given $X_i$.

\subsection{Overview of Proposed Changes}
Our proposed model is an extended seq2seq model in which we employ a hierarchical attention mechanism to create a query-aware attention.
We will show in the experiments that this enables us to not only control the scope of the queries in the session from which we generate the next query, but also to automatically detect session boundaries.
Furthermore, regarding the fact that on average $62\%$ of terms of a query in a session come from the previously submitted queries in the same session~\citep{Sloan:2015} and they are mostly rare terms, we empower our model to access the source words from the session context during decoding by incorporating a copy mechanism. 
A general schema of the proposed model is depicted in Figure~\ref{fig:model}.
In the following, we describe these two functionalities in detail.
\vspace{-10pt}
\subsection{Query-Aware Attention}
We already described the attention mechanism which is in charge of assigning weights to the hidden states of the encoder.
We will call this the word-level encoder and word-level attention mechanism in the remainder of this paper.
We now add a bidirectional RNN query-level encoder, that reads the input sequence $Q = [q_1, q_2,\ldots,q_m]$, whereas $q_j$ is the encoding of query $j$ (blue component in Figure~\ref{fig:model}).
To compute $q_j$ we employ the fact that at word-level the input is linearized by concatenating all queries, each followed by a special token $\langle /q\rangle$.
Let $K \subset \{1, \ldots, n\}$ be the indices of these special tokens, i.e. the inputs $\{x_k | k \in K\}$ corresponds to $\langle /q\rangle$.
Note that $|K| = m$. We annotate the elements in $K$ so that the following holds true: $k_a < k_b \Leftrightarrow a < b$.
We now define the encoding of query $j$ as:
\begin{equation}
q_j = \text{id}(\vec{h}_{k_j})
\end{equation}
with $\text{id}$ being the identity function in our experiments.
However, this function can also be replaced by a small MLP. The query-level encoder is again a bidirectional RNN with the sequence of hidden states of the forward RNN being $[\vec{g}_{1},\vec{g}_{2},\ldots,\vec{g}_{m}]$, where ${g}_j = \text{RNN}(q_j,\vec{g}_{j-1})$, i.e. the query encoding $q_j$ is the input to the query-level encoder. Analog the backward RNN hidden states $[\cev{g}_{m},\cev{g}_{m-1},\ldots, \cev{g}_{1}]$ are computed by $\cev{g}_j = \text{RNN}(q_j,\cev{g}_{j+1})$

The forward and backward states for each time step are again concatenated to create the encoder hidden states $[g_1,g_2,\ldots,g_m]$, in which $g_i=[\vec{g}_i;\cev{g}_i]$.

We now define the query-level attention weights over the query encoding:
\begin{align}
l^q_{t,j} &= \eta(s_{t-1}, g_j, y_{t-1}) \\
a^q_{t,j} &= \frac{\exp(l^q_{t,j})}{\sum_i^n\exp(l^q_{t,i})}
\end{align}

To get the final query-aware attention weights (red component in Figure~\ref{fig:model}) we multiply the word-level attention weight $a^w$ (previously $a$ in Equation \ref{worl_level_att}) with the corresponding 
%
% query-level attention weights:
% \begin{equation}
% a_{t,i} = a^q_{t,j} a^w_{t,i}
% \end{equation}
%  
query-level attention weights and renormalize it as shown below:
\begin{equation}
a_{t,i} = \frac{a^w_{t,i} a^q_{t,j}}{\sum_{i'}^n a^w_{t,i'} a^q_{t,j'}},
\end{equation}
whereas $j$ and $j'$ is the smallest $j$, respectively $j'$, so that $i \leq k_j$, respectively $i' \leq k_j'$, holds true.
Finally, the context vector is computed as stated in Equation \ref{context_vector}.

\vspace{-10pt}
\subsection{Copy and Generate for Decoding}
As already mentioned, we use an output projection layer to compute the conditional distribution over the vocabulary.
This allows us to generate words, that are part of our vocabulary (yellow components in Figure~\ref{fig:model}):
\begin{equation}\label{output_projection_2}
p(y_t|y_{<t}, X, \text{generate}) = f_o(s_t)
\end{equation}
However, a large vocabulary will slow down the training process and increase the model size.
Additionally, some entities or numbers (e.g. telephone numbers) will not be part of even a very big vocabulary.

We therefore employ a copy mechanism (green component in Figure~\ref{fig:model}).
This copy mechanism (also know as Pointer Network~\citep{Vinyals:2015}) computes a probability distribution over the input sequence, not by using a fixed length output projection matrix but by using the hidden states of the encoder:
\begin{align}
l^p_{t,i} &= \eta(s_{t}, h_i) \\
p(y_t=x_i|y_{<t}, X, \text{copy}) &= \frac{\exp(l^p_{t,i})}{\sum_{j=1}^n\exp(l^p_{t,j})}
\end{align}
Note, that this equation is very similar to the attention mechanism.
Instead of using the final weights as attention weights, we use them as probabilities for copying the underlying word.
In order to be able to handle words that have to be generated, i.e. are not in the input, we define $x_0$ to be the unknown token ($\langle UNK\rangle$) and change the equation above as follows:
\begin{align}
l^p_{t,0} &= \eta(s_{t},e(x_0)) \\
l^p_{t,i} &= \eta(s_{t}, h_i) \qquad i > 0 \\
p(y_t=x_i|y_{<t}, X, \text{copy}) &= \frac{\exp(l^p_{t,i})}{\sum_{j=0}^n\exp(l^p_{t,j})}
\end{align}
with $e$ being an embedding function. 
We also include a switch gate, that decides whether the generated or the copied word is used as the final prediction of the decoder (grey component in Figure~\ref{fig:model}). The switch gate has to make a binary decision and is defined as follows:
\begin{align}
p(\text{copy}) &= \sigma(w^T s_t) \\
p(\text{generate}) &= 1 - p(\text{copy})
\end{align}
whereas $w$ is a weight vector and $\sigma$ is the sigmoid function, i.e.:
\begin{equation}
\sigma(x) = \frac{1}{1-\exp(-x)}
\end{equation}
In other words, we have a fully-connected layer with an output size 1 and the sigmoid as the activation function.

During decoding, we perform a beam search of size $b$. At each decoding step $t$ we produce the top $b$ tokens and store them together with their probabilities:
\begin{equation}
\big(y^1_1, p(y^1_1)\big), \ldots, \big(y^b_1, p(y^b_1)\big)
\end{equation}

In the following step we again produce the top $b$ tokens, this time for each of the tokens of the first step. We store them together with the probability of predicting the sequence $y_1, y_2$:
\begin{equation}
\big(y^1_2, p(y^1_2|y^1_1)\big), \ldots, \big(y^b_2, p(y^b_2|y^1_1)\big), \ldots, \big(y^1_2, p(y^1_2|y^b_1)\big), \ldots, \big(y^b_2, p(y^b_2|y^b_1)\big)
\end{equation}

We then drop all but the $b$ most likely decoding paths, which will be use to produce the next token.

\subsection{Multi-Objective Training}
To train our model we define two losses. The loss of the generator is the averaged cross entropy between our predicted probability distribution $p$ and the target probability distribution $q$, i.e. the one-hot encoding of the target token:
\begin{equation}
\text{loss}_\text{generate} = \frac{1}{|V|} H(p,q) = \frac{1}{|V|} \sum_{v \in V}{p_v \log{q_v}}
\end{equation}
with $V$ being the vocabulary and $|V|$ its size.
Similarly we define the loss of the copier as being the cross-entropy averaged over the length of the input:
\begin{equation}
\text{loss}_\text{copy} = \frac{1}{|X|} H(p,q) = \frac{1}{|X|} \sum_{x \in X}{p_x \log{q_x}}
\end{equation}

In order to avoid producing $\langle UNK\rangle$, i.e. the output of copier when the target is not in the source sequence, and $\langle OOV\rangle$, i.e. the output of generator when the target is not in the vocabulary, we introduce a set of rules defining if the switch gate should favor the copier or the generator at a given time step:
\begin{enumerate}
\item target copier is $\langle UNK\rangle$ and target generator is not $\langle OOV\rangle$: the switch gate shall choose generation ($t_\text{switch} = 0$).
\item target copier is not $\langle UNK\rangle$ and target generator is $\langle OOV\rangle$: the switch gate shall choose copying ($t_\text{switch} = 1$).
\item target copier is $\langle UNK\rangle$ and target generator is $\langle OOV\rangle$: the switch gate shall choose generation ($t_\text{switch} = 0$).
\item target copier is not $\langle UNK\rangle$ and target generator is not $\langle OOV\rangle$: the switch gate shall choose copying ($t_\text{switch} = 1$).
\end{enumerate}
While the first two rules are an obvious choice, the last two rules derive from the fact that we want to choose a target label for the switch gate to copy as much as possible from the input and let the generator handle the rest.  Note that the loss of the generator will be $0$ if the target token is $\langle OOV\rangle$ and the generator predicted $\langle OOV\rangle$. The same applies to the copier choosing $\langle UNK\rangle$.
Now, based on these rules we can define the loss of the switch gate to be the following:
\begin{equation}
\text{loss}_\text{switch} = (p(\text{copy}) - t_\text{switch})^2
\end{equation}

During the backward pass of the back-propagation algorithm, we update the parameters of the network with respect to the losses in three separate steps.
In step one, we use the gradient calculated from $\text{loss}_\text{copy}$ to update all parameters of the network, except those just belonging to the switch gate and the generator, by freezing these components (yellow and gray components in Figure~\ref{fig:model}). 
In step two, we propagate the gradients calculated from $\text{loss}_\text{generate}$ to update all parameters of the network except the parameters of the switch gate and the copier (green and gray components in Figure~\ref{fig:model}). 
In the last stage, we update the parameters of the network using the gradients from $\text{loss}_\text{switch}$, while the parameters of copy and generator (green and yellow components in Figure~\ref{fig:model}) are fixed.
We empirically found that updating parameters in separate stages instead of combining losses as one loss leads to better results.

\section{Evaluation}
With the increase of interest in generative models and in particular neural network based models, we believe automatic evaluations have to be studied as part of this research.  

In this paper, we evaluate the performance of our proposed model and the baselines using two paradigms. 
In the first paradigm, similar to the previous studies, we assess the ability of the model as a discriminative algorithm.
While in the second one, we introduce metrics to investigate the quality of the model in terms of generating data as a step toward evaluation of the generative models for the query suggestion task.

\subsection{Evaluation based on Discrimination}
\label{sec:disev}
In this paradigm, similar to  \citep{Sordoni:2015,Santos:2013,Ozertem:2012,Mitra:2015} we evaluate our model as a feature to score the candidate queries and use it within the learning-to-rank (L2R) framework for ranking candidate suggestions. 
For the sake of a fair comparison, we replicate the experimental setups in the paper by ~\citet{Sordoni:2015} by generating candidates using the co-occurrence based model they used and also extract the set of features they employed to be used in an L2R method as the BaseRanker.
We compare the performance of the BaseRanker with a similar ranker that gets the score from our model as an additional feature. 

The score of a query $q = [y_1, \ldots, y_n]$ given the context $X$ is calculated based on the output of the generator, the copier, and the switch gate for each term in $q$ as follows:
\begin{equation}
\label{equ:acg_score}
\begin{aligned}
p(q|X) = \prod_{t=1}^{n} 
~ &\Big( p(\text{generate}|y_{<t}, X)~p(y_t|y_{<t}, X, \text{generate}) \\
~ &+ p(\text{copy}|y_{<t}, X)~p(y_t|y_{<t}, X, \text{copy})\Big)
\end{aligned}  
\end{equation}
We use the mean reciprocal rank (MRR) to measure the quality of the ranking.

\subsection{Evaluation based on Generation}
\label{sec:genev}
Besides the aforementioned evaluation paradigm, we also evaluate the quality of our model's output as a generative model.
Loosely speaking, we try to evaluate how similar the generated query is to the target query taken from ground truth. To do so, we introduce the following metrics.

\mypar{Word Overlap Based Query Similarity.}
As a word overlap based metrics, we consider Position Independent Word Error Rate (denoted as \emph{PER}), which is the 
%minimum number of substitutions, deletions, and insertions
minimum number of word insertions and deletions necessary to transform the generated query into the target query by neglecting the words order, normalized by the length of the target query.

\mypar{Embedding Based Query Similarity.}
Word overlap based metrics are not efficient in assessing the quality of the generative models as they may generate queries that are as effective as the target query without sharing any term with it. For example consider the target query \emph{city hall phone number} and a system that generated query \emph{municipality contact information}, where the score is zero in terms of PER.
To tackle this issue, we can use soft matching metrics. In this paper, we used an embedding-based query similarity metric (denoted as $\text{sim}_\text{emb}$) to assess the semantic matching of the target and generated query. 
We first calculate the query-level embeddings using vector extrema~\citep{Forgues:2014} for which we use pretrained word embeddings (trained on the GoogleNews corpus) and for each dimension
of the word vectors, we take the most extreme value amongst all word vectors in the query and use that value in the query-level embedding.
This approach prioritizes words carrying important semantic information over common ones~\citep{Liu:2017}.
Then, we compute the similarity between the generated query and the target query vectors using the cosine similarity.

% \vspace{-5pt}
\mypar{Retrieval Based Query Similarity.}
In the real world's application of the query suggestion task, the suggested query is going to be submitted to a search engine if it was selected by the user. 
Utilizing this fact, we can evaluate the quality of the suggested query in terms of how similar the retrieved results from the search engine with respect to this query are to the results retrieved using the target query.
To this end, we suggest three different evaluation metrics:

To calculate the first metric, we submit the target query to an external collection of documents and retrieve the ranked list of top documents using a retrieval function and consider this list as the reference list.
We do the same but given the generated query.
Then, we use a ranking similarity metric to calculate the agreement of these two ranked lists (denoted as $\text{sim}_\text{ret}$).
%It could be controversial so I guess we need to elaborate this:
We are not directly measuring if the suggested query is addressing the actual user information need.
But we can reasonably conclude how well we do in terms of helping a user which lacks the ability for query reformulation by suggesting him a query which retrieves results that are similar to those of a well-reformulated query.

To implicitly tackle this issue of not measuring how we address the user information need, we define the second metric.
We consider the ranked list retrieved from the expanded target query using Pseudo Relevance Feedback (PRF) as the reference list. 
It has been shown that PRF usually improves the performance of retrieval~\citep{Dehghani:2016, Harman:2009} by decreasing the vocabulary gap of a query and relevant documents. 
Hence considering its results as the reference list, we can estimate how well the generated query performs in terms of retrieving results from an (in average) better version of the target query, which is more likely to address the user information need (denoted as $\text{sim}_\text{ret}^{+}$).

In order to even better evaluate how good the generated query is addressing the actual user information need, we take all sessions with length $l>2$ from the test data.
We then use the first $\lfloor \nicefrac{l}{2} \rfloor$ queries in the session as the context for generating the next query.
After this, we retrieve the ranked lists of documents for each of the next $\lceil \nicefrac{l}{2} \rceil$ queries in the session and merge them using normalized scoring.
The merged list is used as the reference ranked list and we calculate its agreement with the retrieved results given the generated query (denoted as $\text{sim}_\text{ret}^{++}$). 

The two last proposed metrics are in fact aiming at better evaluation of generated suggestions with respect to the actual user information needs when the user judgments are not available. 

\section{Experiments and Results}
In this section, we first describe the dataset we used for training, validation, and testing as well as our experimental setups. 
Then we present the main results of the model followed by discussions and analysis. 

\subsection{Dataset and Experimental Setups}`
We used AOL search logs~\citep{Pass:2006} as the largest publicly available search log which is also used by similar research works for the evaluation~\citep{Sordoni:2015,Sengstock:2011,Hsu:2013,Bar:2011,Dehghani:2017}. 
This query set contains web queries initiated by 657,426 unique users in the AOL search engine that were sampled from a three-months period from March 1, 2006, to May 31, 2006. 
We preprocessed the data by eliminating non-alphanumeric characters, spelling error correction, and lowercasing. 
Then we segmented the log into sessions, using a simple standard segmentation heuristic, i.e. intervals of at least 30 minutes idle time denotes a session boundary~\citep{Jansen:2007}. 
The final dataset consists of $\sim9$ million queries in $\sim3$ million sessions.
We sorted the sessions time-wise and partitioned them into three parts: the main training set, consisting of 70\% of the sessions which is used to train our proposed model and the baseline methods.
We also created an L2R training set, which includes 20\% of sessions for training the L2R algorithm (just used in the evaluation based on discrimination).
10\% of sessions are used as a test set.

The evaluation based on discrimination is similar to previous research~\citep{Sordoni:2015,Santos:2013}.
Given the anchor query, i.e. the last query in the context, we first select top-20 candidate queries ranked based on the frequency of their co-appearances with the anchor query in the same session in the main training set, as the Most Popular Suggestions (\textbf{MPS}).
We then extract a set of 17 features employed by ~\citet{Sordoni:2015}.
These include:
\begin{enumerate}[leftmargin=*]
\item Features which capture the whole session history, like the score calculated by Query Variable Markov Model as one of the context-aware query suggestion models~\citep{He:2009}, character n-gram similarity between the candidate query and the 10 most recent queries in the context~\citep{Mitra:2015,Shokouhi:2013}, and average Levenshtein distance between the candidate and queries in the context~\citep{Jiang:2014}.
\item Features which only take the most recent query into account, like frequency of the anchor query in the main training set, the number of times the candidate follows the anchor query in the main training set, and Levenshtein distance between the anchor query and the candidate~\citep{Jiang:2014,Ozertem:2012}.
\item Features which characterize the candidate independently, like frequency of the candidate in the main training set and the length of the candidate in terms of the number of words and characters.
\end{enumerate}
When we extracted features for all candidates, we train the LambdaMART as the L2R method by labeling the target query as relevant and all others as non-relevant.
We call this trained model BaseRanker.
We use the BaseRanker in addition to the score from ~\citet{Sordoni:2015} paper as an additional feature, as one of the baselines. 

Besides these baselines, we train the original seq2seq model with attention (explained in Section~\ref{sec:seq2seq_attention}) with the whole context flattened to a sequence as the source sequence and the target query as the target sequence. 
We also add the score from this model to the BaseRanker as an additional feature as one of the baselines.
We also add the score from our model (calculated using \ref{equ:acg_score}) to the base-ranker to assess how it further helps to improve the quality of the ranking of suggestions.  
Besides this, we want to investigate how different part of our model including the query-aware attention and the copy mechanism (CM) individually affect the performance of query suggestion,
So we also evaluate the seq2seq model with query-aware attention only (seq2seq + QaA), and seq2seq model with copy mechanism (seq2seq + CM) only as additional baselines.

In the evaluation based on generation, for the retrieval based metrics, we use ClueWeb09 Category B corpus with over 50 million English documents as the external collection of documents.
As the retrieval function, we use JS-divergence retrieval model with the Dirichlet prior smoothing and as the PRF method, we use RM3~\citep{Abdul:2004}.
We further use the rank-biased overlap (RBO)~\citep{Webber:2010} at level $100$ to measure the agreement of two ranked lists and report the average RBOs over all instances in the test set.

We used TensorFlow~\citep{tang2016:tflearn,tensorflow2015-whitepaper} to implement our model.
The parameters of our model are optimized employing the Adam optimizer~\citep{Kingma:2014} and using the computed gradient of the loss to perform the back-propagation algorithm.
With respective to the main validation set, which is 10\% of the sessions from the main training set we tuned some hyper-parameters of the model using batched GP bandits with an expected improvement acquisition function~\citep{Desautels:2014}. 
In our model, the number of hidden neurons in each of encoders (forward and backward; word-level and query-level) and the decoder were selected from $[64, 128, 256]$.
The initial learning rate and the dropout parameter were selected from $\{10^{-3}, 10^{-5}\}$ and $\{0.0, 0.2, 0.5\}$, respectively.
We considered embedding sizes $\{300, 500\}$ for the input embedding of the word level encoder.
The batch size in our experiments was set to $128$.
We set the vocabulary size to $90k$ (the same for the baselines), and the beam size for the beam search decoding was set to $4$.

It is noteworthy that for all the baseline models we also tuned the hyper-parameters if their optimized value is not reported in the corresponding papers.
% The implementation of the model along with the code for replicating the dataset is open-sourced.\footnote{Link to the repository is removed to preserve the blind review policy.}

\subsection{Main Results}
In this section, we first report the main results of our proposed model in the task of query suggestion compared to the baselines, both in terms of discrimination and generation quality. 

%
\input{table_DisEv}
\input{table_GenEv}
\mypar{Evaluation based on Discrimination.}
First, we evaluate the performance of our model as a discriminative model in the setup explained in Section~\ref{sec:disev}.
Table~\ref{tbl_disev} presents the performance of our model and the baseline models. 
As it can be seen, by adding the score from \acg, we can gain the highest improvement compared to the baseline models. 
This is mainly because of the fact that in the scoring process of candidates, rare terms from the true target query that occur in the candidate query have a chance of getting a high probability in the distribution learned by \acg if they appear in one of the previously submitted queries.

It is interesting that the improvement of just having the hierarchical setup for reading the session history, i.e. the source sequence in both HRED and seq2seq with query-aware attention is not statistically significant with respect to the seq2seq model with a simple attention mechanism (models 4 and 5 compared to model 3). 
On the other hand, when the model is equipped with the copy mechanism, the performance improvement by adding query-aware attention is statistically significant (model 7 compared to model 6). 
This suggests that taking query-level attention into account along with the word-level attention,
%helps in particular when we retain terms from the previous queries in the next suggested query.
is particularly important for the copy mechanism which directly uses the attention weights as output probabilities, but is less important for the decoder itself.

\mypar{Evaluation based on Generation.}
The next set of experiments aims at assessing the quality of the models in terms of generating the next query according to the metrics we introduced in Section~\ref{sec:genev}. 
Table~\ref{tbl_genev} presents the performance of different models. 
We also evaluate the top ranked query from different models used in the evaluation based on discrimination setup, but in terms of evaluation based on generation. 

Among the metrics for evaluation based on generation, PER is a strict metric as it is based on the exact word overlap between generated query and target query. 
Nonetheless, the generated next query by \acg has the lowest error in terms of PER, even lower than the corresponding discriminative model in which the score from \acg is used as one of the features (model 7 compared to model 8). 
The relatively low error rate of model 4 is an indicator that the success of \acg to generate exact words in the target query is mostly reasoned by its ability to copy terms from the context queries. 
It is supported by the statistics that we extracted from our test set, where in average 38\% of terms in the target queries are retained from the previous queries in the same session.

Embedding similarity of the generated and the target queries relaxes the hard assumption in PER by taking words semantic similarity into consideration. 
\acg performs better than all the baseline in terms of generating queries that are semantically similar to the target queries.
Regarding the results of the models we report in this paper, generative approaches lead to queries with higher $\text{sim}_\text{emb}$ scores.
On the other hand, in discriminative approaches are in average more successful in terms of PER. 
This is assumed to be reasoned by the fact that in the generative models in this paper, we explicitly learn representations for words, in a downstream task of learning query suggestions, which helps to better capture the semantic similarity among the words.
During generation, the model tries to generate the semantically most plausible words for the suggested query based on the learned representations.  

Regarding the retrieval based evaluation, with respect to the $\text{sim}_\text{ret}$ metric, BaseRanker which uses \acg as one of its features outperforms all baseline models.
In contrast, for both $\text{sim}_\text{ret}^{+}$ and $\text{sim}_\text{ret}^{++}$ our \acg as a generative model is the best performing method. 
Since PRF implicitly models the retrieval with soft matching, using $\text{sim}_\text{ret}^{+}$ we also consider cases where the results given the generated query do not have an immediate agreement with the results given the target query, but the generated query is semantically related to the target query. 
Also with $\text{sim}_\text{ret}^{++}$, we move one step further by evaluating if the results given the generated query are going to satisfy the user at some point in the search session, even if it is not the case given the immediate target query. 
This explains why we achieve better results using generative approaches with respect to these two metrics.

\subsection{Discussions and Analysis}
In this section, we first investigate the generative ability of our model in a multiple query suggestion setup.
We then study the effect of session length on the performance of \acg. At the end, we will assess the robustness of \acg compared to the baseline in dealing with noisy data.

\mypar{Multiple Query Suggestion.}
Major search engines usually provide more than one suggestion for the users. 
Here, we investigate the quality of the top-10 queries that our model generates. 
To do so, during the decoding, after generating the first suggestion, we ignore the fact that the first suggestion was generated through the beam search, i.e. $\langle /q\rangle$ was generated.
Instead, we continue decoding until the next suggestion is generated. 
We repeat this process to generate 10 suggestions which lead to a sorted list of queries based on their likelihood of generation.
\input{Images/chart_2}

The performance of \acg considering the queries at different ranks, i.e. position in the list of queries, with respect to the different metrics, is illustrated in Figure~\ref{fig:ms}. 

As it is expected, in general, the quality of the generated queries decreases as the rank of the query increases. 
The drops in terms of
$\text{sim}_\text{ret}$, $\text{sim}_\text{ret}^{+}$, and  $\text{sim}_\text{ret}^{++}$ are less dramatic compared to $\text{sim}_\text{emb}$. 

It is also really interesting that the performance of the secondly generated query is relatively low compared to the first query in terms of all metrics except the $\text{sim}_\text{ret}^{++}$, where the score is even slightly higher than the score for the first query. 
Although this boost is not statistically significant. 
But regarding the meaning of the $\text{sim}_\text{ret}^{++}$ metric (Section~\ref{sec:genev}), it seems the generated queries in the second and the third rank are somehow future desirable queries which although they do not match with the immediate user query, are related to the ultimate user information need.

\mypar{Effect of Session Length.}
In order to investigate the effect of session length on the performance of our model, we separate the test set into three categories:
\begin{enumerate}[leftmargin=*]
\item sessions with $2$ queries (short) 66.09\% of the test set
\item sessions with $3-4$ queries (medium) 22.36\% of the test set
\item sessions with $>4$ queries (long) 11.55\% of the test set
\end{enumerate}
We report the performance of our model in terms of both ability of discrimination and generation, compared to the HRED~\citep{Sordoni:2015}, using test sets with short, medium, and long sessions. 
The results are depicted in Figure~\ref{fig:sl}. 

In Figure~\ref{fig:sl-mrr}, we evaluate the performance of our model and the baseline model as an additional feature for the BaseRanker and measure the MRR of the ranked list of suggestions.  
\acg appears to perform best in all test sets and its performance is not only stable across the test sets with different session lengths but also progressive when the length of the sessions increases. 
However, HRED seems to fail for long sessions, which can be the result of topically broad information needs or changes of the search topic. 
This problem is quiet challenging, although our model seems to handle this problem by dynamically attending to the most promising part of the context.  

This is more the case for evaluation based on generation.
However, it has been shown that when the length of a session increases, the percentage of repeating previously-used terms also increases~\citep{Jiang:2014}, which means that the next query is more likely to contain the terms used before when it appears in the latter steps of a session. 
This is where the ability of \acg for copying terms from the context queries kicks in that compensates the performance loss which might occur due to the complexity of the long sessions.
\input{Images/chart_1}

\mypar{Attending the Promising Parts of the Context.}
Analyzing the context of a session to understand which part is useful for query suggestion is considered as one of the most challenging stages of this task.
This is due to the fact that noisy words are common to appear in a given context and this can happen in several circumstances: 
Either when users switch to another search topic quickly after one search session and the assumption of 30 minutes as the idle time duration for detecting session boundaries fails. 
Or when users issue navigational queries during a search session, like e.g. \emph{google}. 
Or in cases where users use non-discriminating words in their query that are no effective indicators for the search intend.

In this situation, a session-aware query suggestion system should be able to handle noisy parts by neglecting them.
To test the ability of our model compared to the baselines on being robust against these noises, we manipulate the sessions in training, test, and validation sets in three ways.
We train and evaluate the query suggestion models on this manipulated data:
\begin{itemize}[leftmargin=*]
    \item \emph{Noise term insertion}:
    In order to assess the robustness of the models against noise terms in the session, we first select the 200 most frequent terms excluding stopwords from the main training set. 
    Then, for each session in the data, we sample from this list with a probability that is proportional to the frequency of the terms and insert the sampled noise term at a random position of a random query in the session.  
    \item \emph{Noise query insertion}:
    In order to assess the robustness of the models against noise queries, similar to the previous case, we extract a list of 100 most frequent queries in the main training set as noise queries. For each session, we sample from the list and insert the noise query at a random position in the session.
    \item \emph{Noise session insertion}:
    In the last case, we aim at assessing the ability of the models in session boundary detection. To do so, for each session, we randomly pick another session from the same user, if there is any, and insert it at the beginning of the session.
\end{itemize}

Table~\ref{tbl_robust} presents the results of seq2seq, HRED, and \acg as generative models in terms of $\text{sim}_\text{emb}$. 
As shown, \acg achieves a significantly better performance in all situations compared to HRED and seq2seq.
In terms of performance loss compared to the non-noisy case (results Table~\ref{tbl_genev}), all models are relatively robust to noise term insertion. 
However, seq2seq is not able to handle noise query and noise session insertion.
Although both HRED and \acg seem to be still robust against noise query insertion, inserting noise sessions considerably affects the performance of HRED, which is also in accordance with the performance drop of HRED in encoding long sessions (Figure~\ref{fig:sl}). 
This is while \acg controls the performance loss in this situation by providing insight into how different parts in the given context contribute to the next query generation. 
In other words, \acg is able to detect the boundary of the session by attending to the promising scope of the context.

\input{table_robust}

\mypar{Error Analysis.}
As discussed and showed in the experiments, the main superiority of the model over the HRED, i.e. the main baseline, is its ability to copy an arbitrary part of the input sequence during decoding. 
We analyzed cases where \acg fails against HRED, in different situations and the main point of failure of \acg is where we remove the spelling correction from the preprocessing. Looking into the output of our model we noticed that it usually fails in cases where the next query in the session is just a spelling correction of the previous queries and the copy mechanism tends to repeat the spelling error. 
For instance, given the query ``\emph{bebefit pediatrician}'', \acg suggests ``bebefit pediatric center'' as the next query, while the obvious reformulation for the given query would be the replacement of ``\emph{bebefit}'' with ``\emph{benefit}''. 
The easiest way to tackle this problem is to have spell error correction as part of the preprocessing. We also consider training the model with a spelling correction dataset that helps to pick generating the correct version of words with errors instead of copying them. We will leave this for future work.

\section{Related Work}
\mypar{Query Suggestion.}
The biggest challenge in the task of query suggestion is to understand the actual user intent for suggesting the next query.
To aim this, several studies leveraged the ``wisdom of crowds'' by mining the session structure and other information in query log data to find alternative queries for suggestions.
For example, \citet{Huang:2003} tried to find within session query pairs co-occurrences and rank suggestions based on the frequency of co-occurrence. \citet{Boldi:2009} took the structure of the sessions into account by building a query-flow graph to estimate how likely a user moves from one query to the next query in the same session. 
Another group of methods tries to find similar queries in log data.
It is assumed that similar queries have larger overlap between their respective clicks and based on this assumption.
E.g., ~\citet{Mei:2008} used a random walk over a bi-partied query-document graph or ~\citet{Baeza:2004} used k-means to find similar queries based on clicked data.
~\citet{He:2009} proposed a session-based method based on Variable Memory Markov model (QVMM) to build a suffix tree to model the user query sequence for query suggestions.
\citet{Santos:2013} and \citet{Ozertem:2012} tried to first extract candidates and then employed learning-to-rank methods for ranking suggestions considering several features (similar to our BaseRanker baseline).
These methods perform poorly for long-tail queries.
To tackle this issue, \citet{Vahabi:2013} proposed to find suggestions for long-tail queries by comparing their search results. 
Another group of research tries to generate new suggestions by leveraging search logs and external resources.
E.g., ~\citet{Szpektor:2011} used WordNet along with a template generation method, or ~\citet{Jain:2011} used a CRF for segmenting queries followed by a machine learning stage to filter out poor suggestions.
Recently, \citet{Sordoni:2015} proposed a context-aware method which uses a hierarchical RNN to encode the session information and generate the sequence of words as the next query.
Their work is the most similar to ours.
We additional introduce the query-aware attention for capturing the hierarchical structure of the session.
We also augment our model with copy mechanism to better model the query reformulation which lead to a significant improvement in the performance.

\mypar{Hierarchical Structure Decoding.}
There are several attempts to make recurrent neural networks able to deal with structured data.
E.g., \citet{Li:2015} proposed a hierarchical auto-encoder to encode and reconstruct multi-sentence paragraphs, taking both word and sentence levels into account.
\citet{Serban:2016} used a similar architecture to \citet{Sordoni:2015} paper in the context of dialogue systems to encode the context of the dialogue at the utterance level.
\citet{Chung:2017} proposed a model encodes the temporal dependencies with different timescales by updating probabilities for different units to capture the
latent hierarchical structure in the sequence. 
\citet{Yang:2016} proposed a two-level attention mechanism which uses the word level attention to lean the sentence level attention to improve the performance of document classification. 
The query-aware attention mechanism we propose in this paper is similar to theirs.
However, we learn the query-level and word level attentions in separate processes and then integrate information from both during decoding.

\mypar{Incorporating Copy Mechanism.}
The idea of incorporating a copy mechanism originally comes from the pointer networks~\citep{Vinyals:2015}, which is in fact a seq2seq model with attention that produces an output sequence consisting of elements from the input sequence.
The pointer network has been used to create hybrid approaches that mix copying (pointing) and generation during decoding in different tasks like neural machine translation~\citep{Gulcehre:2016}, language modeling~\citep{Merity:2016}, and summarization~\citep{Gu:2016,Nallapati:2016,See:2017}. 
Our approach is the close to \citet{Gu:2016} and \citet{See:2017} works but we considered learning to copy, to generate, and to make the decision of copying or generating as separate tasks by using multi-objective learning.

\section{Conclusion and Future Works}
In this paper, we proposed a session-aware query suggestion model by augmenting the seq2seq model with a query-aware attention mechanism to make it able to encode the structure of the session.
We also incorporate a copy mechanism during decoding which helps to model term retention in query reformulation.
Finally, we proposed new metrics to evaluate generative models on the task of query suggestion. 
Our experiments show that our proposed model boost the performance of seq2seq models and outperforms baselines both in terms of discrimination and generation.

For future work, we are going to extend our model to integrate information from clicked documents as additional signals in the process of generating the next query. 
This information also can be used for the evaluation based on generation paradigm.
For evaluation purposes, we can also use the content of documents retrieved using the generated query and estimate how well the generated query addresses the user information need in terms of the ability to provide relevant content.
Another interesting direction is to evaluate our model in a query auto-completion task. This is possible by changing the setting of the model to have a character-based seq2seq model.
Besides encoding the previously submitted queries in the session, we can feed the decoder with the prefix of the new query and generate the most probable sequence of following characters.

% \begin{acks}
% \end{acks}
\bibliographystyle{ACM-Reference-Format}
\bibliography{ref} 

\end{document}

%% file: table_DisEv.tex
% !TEX root = main.tex

\begin{table}[tbp]
\centering
\caption{\label{tbl_disev}Performance of the different methods as discriminative models. $^{(x)}$ indicates that the improvements with respect to the method in row $x$ is statistically significant, at the 0.05 level using the paired two-tailed t-test with Bonferroni correction.}
%\begin{adjustbox}{max width=\columnwidth}
%\begin{tabular}{@{}l@{~~~}c@{~~}c@{~~}c@{~~~}c@{~~}c@{~~}c@{}}
\begin{adjustbox}{max width=0.9\textwidth}
\begin{tabular}{l l l}
\toprule
$\#$ & Model & \textit{MRR}
\\ \midrule
1 & \textbf{MPS} 
& 0.5216
\\ 
2 & \textbf{BaseRanker} 
& 0.5530$^{(1)}$ 
\\
3 & \textbf{BaseRanker + Seq2Seq} 
& 0.5679$^{(1,2)}$ 
\\
4 & \textbf{BaseRanker + HRED~\citep{Sordoni:2015}} 
& 0.5727$^{(1,2)}$ 
\\
5 & \textbf{BaseRanker + (Seq2Seq + QaA)} 
& 0.5744$^{(1,2)}$ 
\\
6 & \textbf{BaseRanker + (Seq2Seq + CM)} 
& 0.5851$^{(1,2,3,4,5)}$
\\
7 & \textbf{BaseRanker + \acg} 
& 0.5941$^{(1,2,3,4,5,6)}$
\\ \bottomrule
\end{tabular}
\end{adjustbox}
\end{table}

%% file: table_GenEv.tex
\newcommand{\ps}{$^\blacktriangleup$}
\newcommand{\ns}{$^\smalltriangledown$}
\newcommand{\fs}{$^{~}$}
\newcommand{\mypm}{\mathbin{\smash{% 
    \raisebox{0.35ex}{%
            $\underset{\raisebox{0.5ex}{$\smash -$}}{\smash+}$%
            }%
        }%
    }%
}

\begin{table*}[tbp]
\centering
\caption{\label{tbl_genev}Performance of the different methods as generative models.} 
% *** indicates that the improvements with respect to are statistically significant, at the 0.05 level using the paired two-tailed t-test.
%\begin{adjustbox}{max width=\columnwidth}
%\begin{tabular}{@{}l@{~~~}c@{~~}c@{~~}c@{~~~}c@{~~}c@{~~}c@{}}
\begin{adjustbox}{max width=0.9\textwidth}
\begin{tabular}{l l c c  c c c}
\toprule
\multirow{2}{*}{$\#$} &
\multirow{2}{*}{\textbf{Method}} &
\multicolumn{1}{c}{\textbf{Overlap Based}} & \multicolumn{1}{c}{\textbf{Embedding Based}} &
\multicolumn{3}{c}{\textbf{Retrieval Based}}

\\ \cmidrule(lr){3-3} \cmidrule(lr){4-4} \cmidrule(lr){5-7}
& & \textit{PER (\%)} & \textit{$\text{sim}_\text{emb}$}  & \textit{$\text{sim}_\text{ret}$} & \textit{$\text{sim}_\text{ret}^+$} & \textit{$\text{sim}_\text{ret}^{++}$}
\\ \midrule
1 & \textbf{seq2seq} 
& 84.11 \tiny{($\mypm$ 6.3)} & 0.5170 \tiny{($\mypm$ 0.003)} &  
0.1630 \tiny{($\mypm$ 0.008)}  & 0.2424 \tiny{($\mypm$ 0.009)} & 0.1955 \tiny{($\mypm$ 0.008)} 
\\
2 & \textbf{BaseRanker + seq2seq (top-1)} 
& 72.23 \tiny{($\mypm$ 8.1)} & 0.5019 \tiny{($\mypm$ 0.006)} &   
0.4375 \tiny{($\mypm$ 0.009)}  & 0.3751 \tiny{($\mypm$ 0.008)} & 0.3916 \tiny{($\mypm$ 0.008)} 
\\
3 & \textbf{seqsSeq + QaA} 
& 80.90 \tiny{($\mypm$ 5.0)} & 0.5517 \tiny{($\mypm$ 0.004)} &   
0.2012 \tiny{($\mypm$ 0.009)}  & 0.2916 \tiny{($\mypm$ 0.008)} & 0.2330 \tiny{($\mypm$ 0.008)} 
\\
4 & \textbf{seq2seq + CM} 
& 71.16 \tiny{($\mypm$ 3.5)} & 0.6119 \tiny{($\mypm$ 0.003)} &   
0.3563 \tiny{($\mypm$ 0.009)}  & 0.4173 \tiny{($\mypm$ 0.009)} & 0.3950 \tiny{($\mypm$ 0.008)} 
\\
\midrule
5 & \textbf{HRED~\citep{Sordoni:2015}} 
& 81.50 \tiny{($\mypm$ 4.9)} & 0.5455 \tiny{($\mypm$ 0.004)} &   
0.2667 \tiny{($\mypm$ 0.008)}  & 0.3250 \tiny{($\mypm$ 0.009)} & 0.3443 \tiny{($\mypm$ 0.007)}  
\\
6 & \textbf{BaseRanker + HRED~\citep{Sordoni:2015} (top-1)} 
& 72.36 \tiny{($\mypm$ 7.3)} &  0.5200 \tiny{($\mypm$ 0.004)} &   
0.4504 \tiny{($\mypm$ 0.009)}  & 0.3812 \tiny{($\mypm$ 0.009)} & 0.4091 \tiny{($\mypm$ 0.007)} 
\\
\midrule
7 & \textbf{\acg} 
& \textbf{68.03} \tiny{($\mypm$ 3.6)} & \textbf{0.6473} \tiny{($\mypm$ 0.004)} &   
0.3612 \tiny{($\mypm$ 0.008)}  & \textbf{0.4366} \tiny{($\mypm$ 0.009)} & \textbf{0.4315} \tiny{($\mypm$ 0.008)} 
\\
8 & \textbf{BaseRanker + \acg (top-1)} 
& 70.66 \tiny{($\mypm$ 7.1)} & 0.5196 \tiny{($\mypm$ 0.004)}  &   
\textbf{0.4594} \tiny{($\mypm$ 0.008)}  & 0.3927 \tiny{($\mypm$ 0.009)} & 0.4111 \tiny{($\mypm$ 0.007)}  
\\ 
\bottomrule
\end{tabular}
\end{adjustbox}
\end{table*}

%% file: Images/chart_2.tex
\definecolor{sRed}{HTML}{C11B17}
\definecolor{sMagenta}{HTML}{d33682}
\definecolor{sViolet}{HTML}{6c71c4}
\definecolor{sGreen}{HTML}{859900}
\definecolor{sYellow}{HTML}{b58900}
\definecolor{sOrang}{HTML}{cb4b16}
\definecolor{sBlue}{HTML}{268bd2}
\definecolor{sCyan}{HTML}{2aa198}

\pgfplotstableread{
rank	emb	ret	retp	retpp
1	0.6473	0.3612	0.4366	0.4315
2	0.5861	0.324	0.417	0.4405
3	0.5612	0.311	0.4099	0.4298
4	0.5518	0.307	0.4036	0.4106
5	0.5411	0.2911	0.391	0.3788
6	0.5221	0.2711	0.3819	0.3455
7	0.5081	0.268	0.3744	0.336
8	0.4922	0.2512	0.3711	0.3293
9	0.4763	0.249	0.352	0.3201
10	0.4611	0.2411	0.3118	0.295
}\tableone

\definecolor{b}{HTML}{4981CE}
\definecolor{g}{HTML}{859C27}
\definecolor{r}{HTML}{B22222}
\definecolor{o}{HTML}{FF6600}

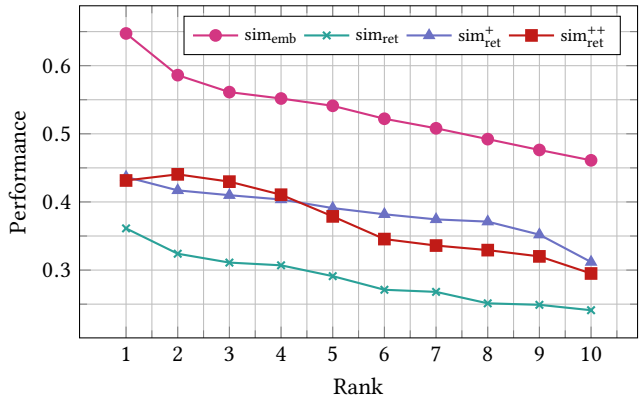
\begin{figure}[!t]
\centering
\begin{tikzpicture}
\begin{axis}[
%xmode=normal,
width= 9cm, % 1*\columnwidth, %\textwidth,
height=6cm,
%enlarge y limits=0.2,
%ymajorgrids,
grid=both,
minor tick num=1,
ylabel={Performance},
xlabel={Rank},
xtick=data,
ytick = {0,0.10,0.20,0.30,0.40,0.50,0.60},
xtick={1,2,3,4,5,6,7,8,9,10},
xticklabels={1,2,3,4,5,6,7,8,9,10},
legend style={at={(0.97,0.97),font=\fontsize{7}{8}}
  ,legend columns=-1
  },
]
\addplot [thick, sMagenta, mark=*] table[x index=0, y index=1]{\tableone};

\addplot [thick, sCyan, mark=x] table[x index=0,y index=2] {\tableone};

\addplot [thick, sViolet, mark=triangle*] table[x index=0,y index=3] {\tableone};

\addplot [thick, sRed, mark=square*] table[x index=0,y index=4] {\tableone};

\legend{$\text{sim}_\text{emb}$,$\text{sim}_\text{ret}$, $\text{sim}_\text{ret}^{+}$, $\text{sim}_\text{ret}^{++}$}

\end{axis}
\end{tikzpicture}
\caption{\label{fig:ms}Performance of the generated queries at different ranks.}
\end{figure}

%% file: Images/chart_1.tex
\definecolor{sRed}{HTML}{C11B17}
\definecolor{sMagenta}{HTML}{d33682}
\definecolor{sViolet}{HTML}{6c71c4}
\definecolor{sGreen}{HTML}{859900}
\definecolor{sYellow}{HTML}{b58900}
\definecolor{sOrang}{HTML}{cb4b16}
\definecolor{sBlue}{HTML}{268bd2}
\definecolor{sCyan}{HTML}{2aa198}

\pgfplotstableread{
RMM	HRED	acg
1   0.5583	0.5861
2   0.6268  0.5925
3	0.5341  0.6040
}\dataone

\pgfplotstableread{
Sim	HRED	acg
1   0.3353	0.4282
2   0.3224  0.4512
3	0.2716  0.4561
}\datatwo

\begin{figure}[t]
\centering
\begin{subfigure}{0.49\columnwidth}
\centering
\begin{tikzpicture}
\pgfkeys{
    % /pgf/number format/precision=2, 
    % /pgf/number format/fixed zerofill=true,
    /pgf/number format/fixed,
}
\begin{axis}[
    width= 5cm, %\textwidth,
    height=4cm, %5cm,
    enlarge y limits=0.05,
    enlarge x limits=0.2,
    ymajorgrids,
    minor tick num=0,
    ybar=0.2pt,
    bar width= 8pt,
    xtick=data,
    xticklabel style = {font=\fontsize{6}{7}\selectfont},
    xticklabels = {small, medium, large},
    ymin=0.4, 
    ymax=0.8,
    ytick = {0.50,0.60,0.70},
    label style = {font=\fontsize{6}{7}\selectfont, yshift=0.5ex},
    anchor=north,
    xlabel={Session length},
    ylabel={MRR},
    area legend,
    legend style={legend image post style={xscale=0.3}, at={(0.49,0.97),font=\fontsize{4}{5}\selectfont},
    legend columns=-1
    },
    nodes near coords,
    every node near coord/.append style={font=\fontsize{4}{5}\selectfont, rotate=90, anchor=west, /pgf/number format/.cd,fixed zerofill,precision=4},
    tick label style={font=\tiny},
    y label style={at={(axis description cs:-0.08,.5)}},
    ]
    \addplot[fill=sOrang, draw=sOrang, pattern color = sOrang, pattern = north east lines
    ] table[x index=0,y index=1] \dataone; 
    
    \addplot[fill=sCyan, draw=sCyan, pattern color = sCyan, 
    pattern = north west lines
    ] table[x index=0,y index=2] \dataone;
    \legend{HRED,\acg}
\end{axis}
\end{tikzpicture}
\caption{\fontsize{6.5}{7.5}\selectfont{Evaluation based on Discrimination}\label{fig:sl-mrr}}
\end{subfigure}
\hfill
\begin{subfigure}{0.49\columnwidth}
\centering
\begin{tikzpicture}
\pgfkeys{
    % /pgf/number format/precision=2, 
    % /pgf/number format/fixed zerofill=true,
    /pgf/number format/fixed,
}
\begin{axis}[
    width= 5cm, %\textwidth,
    height=4cm, %5cm,
    enlarge y limits=0.05,
    enlarge x limits=0.2,
    ymajorgrids,
    minor tick num=0,
    ybar=0.2pt,
    bar width= 8pt,
    xtick=data,
    xticklabel style = {font=\fontsize{6}{7}\selectfont},
    xticklabels = {small, medium, large},
    ymin=0.4, 
    ymax=0.8,
    ytick = {0.50,0.60,0.70},
    label style = {font=\fontsize{6}{7}\selectfont, yshift=0.5ex},
    anchor=north,
    xlabel={Session length},
    ylabel={$\text{sim}_\text{ret}^{+}$},
    area legend,
    ymin=0.1, 
    ymax=0.7,
    ytick = {0.20, 0.30, 0.40, 0.50,  0.60},
    legend style={legend image post style={xscale=0.3}, at={(0.49,0.97),font=\fontsize{4}{5}\selectfont},
    legend columns=-1
    },
    nodes near coords,
    every node near coord/.append style={font=\fontsize{4}{5}\selectfont, rotate=90, anchor=west, /pgf/number format/.cd,fixed zerofill,precision=4},
    tick label style={font=\tiny},
    y label style={at={(axis description cs:-0.08,.5)}},
    ]
    \addplot[fill=sOrang, draw=sOrang, pattern color = sOrang, pattern = north east lines
    ] table[x index=0,y index=1] \datatwo; 
    
    \addplot[fill=sCyan, draw=sCyan, pattern color = sCyan, 
    pattern = north west lines
    ] table[x index=0,y index=2] \datatwo;
    \legend{HRED,\acg}
\end{axis}
\end{tikzpicture}
\caption{\fontsize{6.5}{7.5}\selectfont{Evaluation based on Generative}\label{fig:sl-srp}}
\end{subfigure}
\caption{\label{fig:sl} Performance of \acg compared to HRED on sessions with different lengths.}
\end{figure}
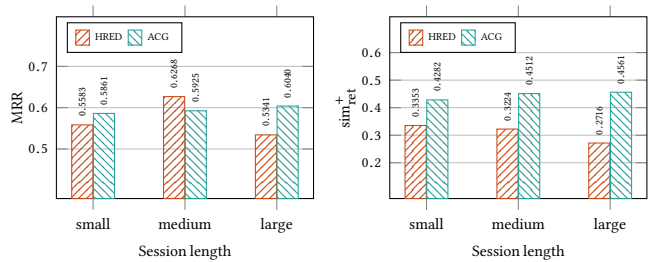

%% file: table_robust.tex
\newcommand{\multilinecell}[2][c]{%
\begin{tabular}[#1]{@{}c@{}}#2\end{tabular}}

\begin{table}[tbp]
\centering
\caption{\label{tbl_robust}Performance (and performance loss) of the different methods as generative models on noisy data, in terms of \textit{$\text{sim}_\text{emb}$}.}
%\begin{adjustbox}{max width=\columnwidth}
%\begin{tabular}{@{}l@{~~~}c@{~~}c@{~~}c@{~~~}c@{~~}c@{~~}c@{}}
\begin{adjustbox}{max width=0.9\textwidth}
\begin{tabular}{l c c c}
\toprule
\textbf{Method} &
\textbf{\multilinecell{Noise term \\ insertion}} & 
\textbf{\multilinecell{Noise query \\ insertion}} &
\textbf{\multilinecell{Noise session \\ insertion}}
\\ \midrule
\textbf{seq2seq} 
& \multilinecell{0.4973  \small{(-3.8\%)}} & \multilinecell{0.4419  \small{(-14.5\%)}} & \multilinecell{0.3969  \small{(-23.2\%)}}
\\
\textbf{HRED~\citep{Sordoni:2015}} 
& \multilinecell{0.5380  \small{(-1.4\%)}} & \multilinecell{0.5140 \small{(-5.8\%)}} & \multilinecell{0.4505  \small{(-17.4\%)}}
\\
\textbf{\acg} 
& \multilinecell{0.6366 \small{(-1.6\%)}} & \multilinecell{0.6019 \small{(-7.0\%)}} & \multilinecell{0.5878  \small{(-9.1\%)}}
\\
\bottomrule
\end{tabular}
\end{adjustbox}
\end{table}